\documentclass[groupedaddress,reprint,amsmath,amssymb,aps,pra]{revtex4-2}
\usepackage{graphicx}
\usepackage{dcolumn}
\usepackage[colorlinks=true, linkcolor=blue, citecolor=blue, urlcolor=blue]{hyperref}
\begin{document}
\title{Bloch Oscillation and Landau-Zener Tunneling of a Periodically Kicked Dirac Particle}
\author{Bin Sun}
\author{Shaowen Lan}
\author{Yi Cao}
\author{Jie Liu}
\email{jliu@gscaep.ac.cn}
\affiliation{Graduate School, China Academy of Engineering Physics, Beijing 100193, China}
\begin{abstract}
    We investigate the dynamics of a relativistic spin-$\frac{1}{2}$ particle governed by a one-dimensional time-periodic kicking Dirac equation. We observe distinct oscillatory behavior in the momentum space and quantum tunneling in the vicinity of zero momentum, which is found to be equivalent to the Bloch oscillations and Landau-Zener tunneling, i.e., Bloch-Landau-Zener (BLZ) dynamics in tilted bipartite lattices. Using the Floquet formalism, we derive an effective Hamiltonian that can accurately predict the oscillation period and amplitude. The tunneling probability has also been determined analytically. Our analysis extends to the influence of various parameters on dynamic behavior. We also discuss how relativistic effects and spin degrees of freedom impact quantum systems' transport properties and localization phenomena.
\end{abstract}
\maketitle
\section{Introduction}\label{sec:1}
The study of regular and chaotic motion in Floquet dynamical systems has been extensively explored in both theoretical and experimental contexts, revealing such as many-body dynamical localization \cite{abaninRecentProgressManybody2017,abaninColloquiumManybodyLocalization2019}, topological phase transitions \cite{rudnerBandStructureEngineering2020,harperTopologyBrokenSymmetry2020}, and quantum chaos \cite{izrailevSimpleModelsQuantum1990,saifClassicalQuantumChaos2005}. A well-established model for illustrating these effects is the quantum kicked rotor (QKR) \cite{santhanamQuantumKickedRotor2022,casatiStochasticBehaviorClassical1979,izrailevQuantumResonanceRotator1980,chirikovQuantumChaosLocalization1988}, which was introduced as a quantum analog of the classical kicked rotor (KR) \cite{chirikovUniversalInstabilityManydimensional1979} to investigate quantum chaos and has been experimentally realized \cite{mooreAtomOpticsRealization1995,ammannQuantumDeltaKickedRotor1998}. One of the most prominent phenomena observed in this system is dynamical localization \cite{casatiStochasticBehaviorClassical1979}, which is closely related to Anderson localization \cite{andersonAbsenceDiffusionCertain1958} in solid-state physics. The QKR model can be directly mapped to the Anderson-type lattice model with quasi-disorder \cite{fishmanChaosQuantumRecurrences1982}.

Despite significant progress in understanding classical chaos and its quantum manifestations, the majority of studies have focused on non-relativistic systems \cite{russomannoChaosSubdiffusionInfiniterange2021,zhaoQuantizationOutoftimeorderedCorrelators2022,zhaoScalingLawsOutoftimeorder2023,tomsovicControllingQuantumChaos2023,zhaoPhaseModulationDirected2024}. An emerging area of interest is the influence of relativistic effects on dynamical behavior \cite{chernikovChaosRelativisticGeneralization1989,kimRelativisticChaosDriven1995,huangRelativisticQuantumChaos2018}, especially about analogies to real-space lattice phenomena in solid-state physics. Matrasulov \textit{et al.} \cite{matrasulovRelativisticKickedRotor2005} explored both classical and quantum dynamics of the spinless relativistic kicked rotor, examining the quantum suppression effects as the system approaches the relativistic limit. Zhao \textit{et al.} \cite{zhaoQuantumClassicalSuperballistic2014} demonstrated that both quantum and classical super-ballistic transport could occur in a relativistic kicked rotor system and discussed the leakage of quantum states from dynamical localization. Rozenbaum \textit{et al.} \cite{rozenbaumDynamicalLocalizationCoupled2017} investigated dynamical localization in coupled relativistic quantum kicked rotors, showing that localization can persist despite many-body interactions.

In this paper, we investigate a spin-$\frac{1}{2}$ relativistic variant of the well-known QKR model \cite{zhaoQuantumClassicalSuperballistic2014,rozenbaumDynamicalLocalizationCoupled2017} (i.e., spin-$\frac{1}{2}$ RQKR) and attempt to explore its dynamical evolution in the momentum space. This kind of periodically driven Dirac system has attracted some experimental interests recently due to advancements in the techniques of the quantum simulations of relativistic particles \cite{gerritsmaQuantumSimulationDirac2010,zhangRelativisticQuantumEffects2012,silvaOpticalSimulationFree2019}. In this spin-$\frac{1}{2}$ RQKR system, in contrast to the conventional QKR model, we find an interesting oscillatory behavior and quantum tunneling in the vicinity of zero momentum, analogous to the Bloch-Landau-Zener (BLZ) dynamics in tilted bipartite lattices~\cite{dreisowBlochZenerOscillations,longhiKleinTunneling}. We also thoroughly discuss how the relativistic effects and spin degrees of freedom influence the quantum localization phenomena. This paper is organized as follows. In Sec.~\ref{sec:2}, we introduce the spin-$\frac{1}{2}$ RQKR model. In Sec.~\ref{sec:3}, we present numerical simulations and theoretical results. We also investigate how relativistic effects and spin degrees of freedom influence the dynamical behavior. Finally, Sec.~\ref{sec:4} concludes this paper.
\section{Model}\label{sec:2}
The dimensionless Hamiltonian of the spin-$\frac{1}{2}$ RQKR is given by \cite{zhaoQuantumClassicalSuperballistic2014,rozenbaumDynamicalLocalizationCoupled2017}
\begin{equation}\label{eq:1}
    \hat{H}=2\pi\alpha\hat{p}\sigma_x+M\sigma_z+K\cos\theta\sum_{n=-\infty}^{+\infty}\delta(t-nT).
\end{equation}
In this model, $2\pi\alpha$ represents the effective speed of light, and $M$ is the particle's mass. The matrices $\sigma_x$ and $\sigma_z$ are the Pauli matrices. $\hat{p}=-i\frac{\partial}{\partial\theta}$ is the one dimensional momentum operator. The term $K\cos\theta$ represents the kicking potential, where $K$ denotes the kicking strength and $\theta$ is confined in $[-\pi,\pi)$. This system is periodic in time, i.e., $\hat{H}(t+T)=\hat{H}(t)$.

The dynamical evolution of the wavefunction $\Psi(t)$ over one period $T$ can be described by a unitary Floquet operator, i.e., $\Psi(t+T)=\hat{F}(T)\Psi(t)$. The explicit expression of the Floquet operator takes the form of $\hat{F}(T)=\hat{\mathcal{T}}\exp\left[-i\int_0^T\hat{H}\mathrm{d}t\right]=\exp\left(-iK\cos\theta\right)\exp\left[-i\left(2\pi\alpha\hat{p}\sigma_x+M\sigma_z\right)T\right]$, where $\hat{\mathcal{T}}$ is the time-ordering operator, and the left term in the last equation represents the sudden perturbation due to each kick and the right term corresponds to free evolution from $t$ to $t+T$.
\section{Numerical Results and Theoretical Analysis}\label{sec:3}
\subsection{Numerical Results of the Bloch-Landau-Zener dynamics}\label{sec:3A}
\begin{figure}[t]
    \centering
    \includegraphics[width=0.45\textwidth]{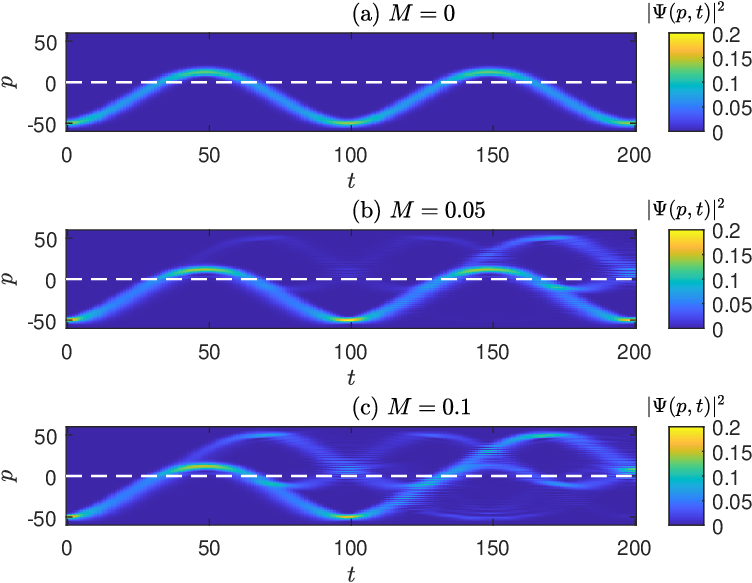}
    \caption{Dynamical evolution of wave packets in momentum space and the white dashed lines indicate line $p=0$. (a) For $M = 0$, the wave packet exhibits perfect oscillations with a period approximately equal to 100. (b) For $M = 0.05$, oscillations are sustained in the first period, but the wave packet splits into two branches when it crosses $p = 0$ in the second oscillation period. (c) For $M = 0.1$, the wavepacket splittings become more pronounced even in the first oscillation period.}
    \label{fig:1}
\end{figure}
We first numerically calculate the time evolution of a spin-$\frac{1}{2}$ Dirac particle. The initial state is set as $\Psi(p,t=0)\sim\exp\left[-\frac{(p-p_0)^2}{2\Delta_p^2}\right]\otimes\chi$, where $p_0$ represents the initial momentum, $\Delta_p$ is the width of the wave packet, and $\chi$ is the spin state. We evaluate the Floquet operator to calculate the dynamical evolution of the particle by switching from the coordinate representation to the momentum representation back and forth. In our numerical calculations, the number of the eigenbasis is chosen to be $2^{12}$, and the numerical convergence has been checked by doubling the number of the eigenbasis.

We choose the typical parameters of the system and initial states as follows: $p_0 = -50$, $\Delta_p = 4$, $\chi = \frac{1}{\sqrt{2}}[1, 1]^T$, $K = 2$, $\alpha = 0.01$ and $T=1$. And our main results are presented in Fig.~\ref{fig:1}.

For the relatively small particle mass situation, for instance, $M=0$ in Fig.~\ref{fig:1}(a), we find that the system exhibits a distinct oscillatory behavior. This phenomenon is different from dynamical localization \cite{casatiStochasticBehaviorClassical1979} found in conventional QKR akin to Anderson localization in quasi-disordered lattices \cite{andersonAbsenceDiffusionCertain1958}. In addition, the oscillation period is around 100, which is significantly longer than the kick period $T = 1$. This kind of spontaneous symmetry breaking associated with the time variable might be related to the concept of time crystal raised recently \cite{wilczekTimeCrystal}. In calculating the momentum spread $\langle p^2\rangle$ \cite{zhaoQuantumClassicalSuperballistic2014,rozenbaumDynamicalLocalizationCoupled2017} in a similar relativistic model, analogous oscillatory behavior is also observed.

With increasing the parameter of the particle mass, as shown in Figs.~\ref{fig:1}(b) and (c), we find that the oscillating wave packet splits into two branches at each time when it crosses the line of $p = 0$. This kind of splitting will eventually break the oscillation of the wave packet, as shown in Fig.~\ref{fig:1}(c), with a larger particle mass ($M = 0.1$).
\subsection{Bloch Oscillation in Momentum Space}\label{sec:3B}
\begin{figure}[t]
    \includegraphics[width=0.45\textwidth]{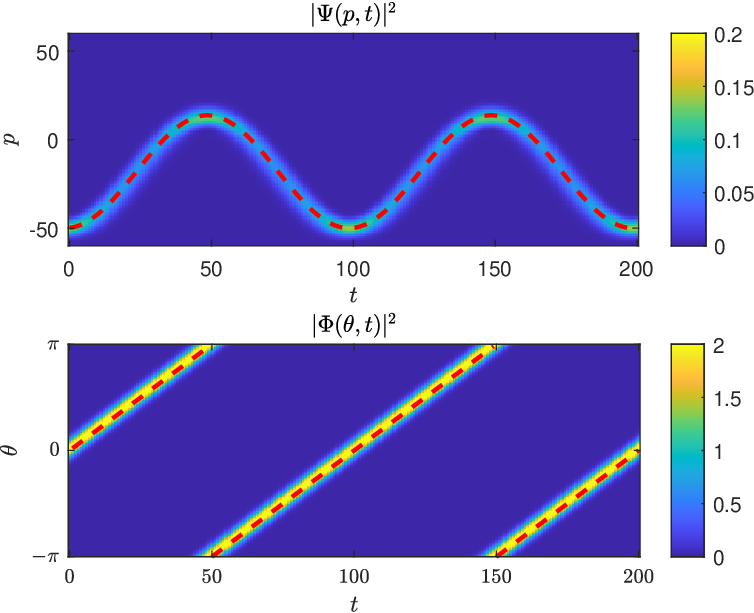}
    \caption{Trajectories of the wave packet center evolution in momentum and coordinate spaces. The red dashed lines represent the theoretical predictions from Eq.~(\ref{eq:4}), which align well with the numerical simulations of wave packet evolution governed by Eq.~(\ref{eq:2}). Parameters are identical to those in Fig.~\ref{fig:1}(a).}
    \label{fig:2}
\end{figure}
In the case of zero mass ($M = 0$), the freedom degrees of the momentum and spin decouple. Corresponding to each eigenspinor of $\sigma_x$ (i.e., $\frac{1}{\sqrt{2}}[1,\pm1]^T$), the Hamiltonian in Eq.~(\ref{eq:1}) reduces to:
\begin{equation}\label{eq:2}
    H_M = \pm 2\pi\alpha\hat{p} + K\cos\theta \sum_{n=-\infty}^{+\infty}\delta(t - nT).
\end{equation}
In the following discussions, the $+$ sign in the above equation is selected since the initial spin state $\chi = \frac{1}{\sqrt{2}}[1, 1]^T$ is used in our simulations.

These Hamiltonians are referred to as the Maryland model \cite{grempelLocalizationIncommensuratePotential1982,longhiMarylandModelOptical2021}. Initially introduced by Grampel \textit{et al.} \cite{grempelLocalizationIncommensuratePotential1982,prangeSolvableModelQuantum1984} and later studied by Berry \cite{berryIncommensurabilityExactlysolubleQuantal1984} and Simon \cite{simonAlmostPeriodicSchrodinger1985}, the Maryland model was employed to investigate Anderson localization analytically. Their work examines how the rationality or irrationality of $\alpha$ influences energy diffusion, specifically the time dependence of $\langle p^2 \rangle$ growth. In Ref.~\cite{berryIncommensurabilityExactlysolubleQuantal1984}, Berry computed the dynamical evolution for a delta-function initial state, which leads to a breathing mode, unlike the oscillating mode observed in Fig.~\ref{fig:1} for a wave packet initial state \cite{hartmannDynamicsBlochOscillations2004}. In this study, we investigate the dynamical evolution of a Gaussian wave packet governed by Eq.~(\ref{eq:2}), using the Floquet formalism \cite{shirleySolutionSchrodingerEquation1965,yangEffectsPeriodicallyKicked2024}.

The Floquet operator $\hat{F}(T)$ of the periodically driven Hamiltonian (\ref{eq:2}) is given by,
\begin{equation}\label{eq:3}
    \hat{F}(T)=\exp\left(-iK\cos\theta\right)\exp\left(-i2\pi\alpha T\hat{p}\right). \\
\end{equation}
According to the Baker-Campbell-Hausdorff formula $e^{\hat{A}}e^{\hat{B}}=e^{\hat{C}}$ with $\hat{C}=\hat{A}+\hat{B}+\frac{1}{2}[\hat{A},\hat{B}]+\frac{1}{12}\left([\hat{A},[\hat{A},\hat{B}]]+[[\hat{A},\hat{B}],\hat{B}]\right)+\cdots$, this Floquet operator $\hat{F}(T)$ can be effectively modeled as $\exp(-i H_M^\mathrm{eff} T)$, where the effective Hamiltonian $H_M^\mathrm{eff}$ is given by:
\begin{equation}\label{eq:4}
    \begin{split}
        H_M^\mathrm{eff} & = 2\pi\alpha \hat{p} + \left(\frac{K}{T}+\frac{\pi^2\alpha^2K}{3}\right)\cos\theta-\pi\alpha K\sin\theta \\&=-i2\pi\alpha\frac{\partial}{\partial\theta}+A\cos(\theta+\delta_s),
    \end{split}
\end{equation}
with $A=K\sqrt{\left(1/T+\pi^2\alpha^2/3\right)^2+(\pi\alpha)^2}$ the amplitude, $\delta_s=\arctan\left[\pi\alpha\big{/}\left(1/T+\pi^2\alpha^2/3\right)\right]$ the phase shift.

This effective Hamiltonian $H_M^\mathrm{eff}$ resembles the form of the quasi-momentum ($\kappa$) Hamiltonian that describes an electron in a one-dimensional periodic potential under the influence of a static force $F$, $H(\kappa)=iF\frac{\mathrm{d}}{\mathrm{d}\kappa}-\frac{\Delta}{2}\cos(\kappa d)$ with $\Delta$ the bandwidth and $d$ the lattice length. In the field of static force, the electron will exhibit an interesting periodic oscillation in the position space, termed the celebrated Bloch oscillation \cite{blochUeberQuantenmechanikElektronen1929,zenerTheoryElectricalBreakdown1934,hartmannDynamicsBlochOscillations2004,breidBlochZenerOscillations2006,flossAndersonWallBloch2014}.

For comparison, we can readily derive the classical motion equations for the wave packet center as $\dot{\theta}_c = 2\pi \alpha$, $\dot{p}_c = A\sin(\theta+\delta_s)$. Considering $(p_0, \theta_0) = (p_0, 0)$ at the initial moment, the explicit solution of these equations is given by:
\begin{equation}\label{eq:5}
    \begin{split}
        \theta_c(t) & = 2\pi \alpha t,                                                                         \\
        p_c(t)      & = p_0 +\frac{A}{2\pi\alpha}\cos\delta_s-\frac{A}{2\pi\alpha}\cos(2\pi\alpha t+\delta_s).
    \end{split}
\end{equation}
We predict a periodic oscillation in the momentum space from Eq.~(\ref{eq:5}). The period of the Bloch oscillation in momentum space is given by $T_B=1/\alpha$, which is inversely proportional to $\alpha$ and independent of all other system parameters. The oscillation amplitude $A/2\pi\alpha$, which explicitly depends on the kicking strength $K$, kicking period $T$, as well as the effective speed of light $2\pi\alpha$. As illustrated in Fig.~\ref{fig:2}, the above simple theoretical predictions align well with our numerical simulations.

\subsection{Multiple-Passage Landau-Zener Tunneling}\label{sec:3C}
\begin{figure}[t]
    \includegraphics[width=0.45\textwidth]{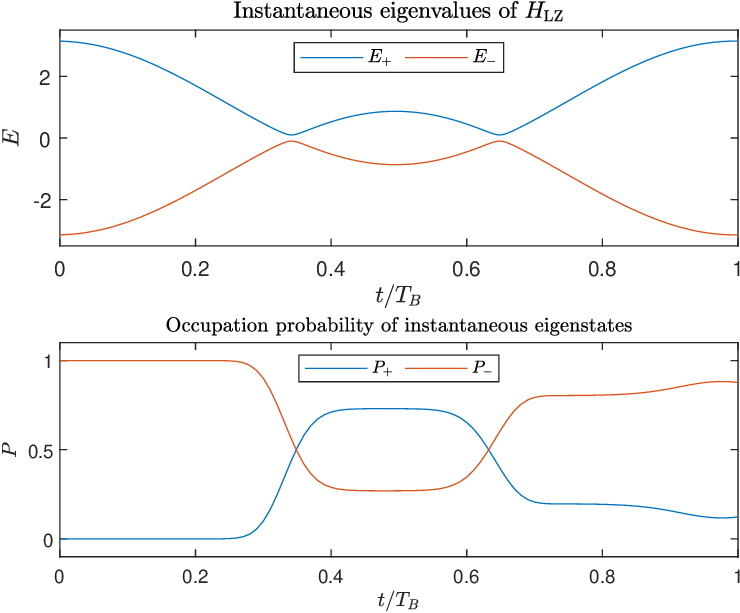}
    \caption{Instantaneous eigenvalues of $H_\mathrm{LZ}$ and occupation probability of instantaneous eigenstates over one Bloch period. Upper: The instantaneous eigenvalues for $H_\mathrm{LZ}$, $E_\pm=\pm\sqrt{(2\pi\alpha p_c(t))^2+M^2}$, correspond to the eigenergies at the wave packet center. Two anti-crossings correspond to the wave packet passing through the $p=0$ line during one Bloch oscillation. Lower: The time evolution of the occupation probability $P_\pm$ for each energy level, driven by Landau-Zener tunneling, as the wave packet passes through the anti-crossing. All parameters are identical to those in Fig.~\ref{fig:1}(c).}
    \label{fig:3}
\end{figure}
As illustrated in Figs.~\ref{fig:1}(b)(c),  the wave packets split into two branches when they cross the line of $p=0$ each time. The partition of the density distribution of the split wave packets relies on the particle mass $M$. This wave packet splitting can be attributed to the Landau-Zener tunneling \cite{landauZurTheorieEnergieubertragung1932,zenerNonadiabaticCrossingEnergy1932,grifoniDrivenQuantumTunneling1998,liuTheoryNonlinearLandauZener2002,shevchenkoLandauZenerStuckelberg2010,wangAdiabaticityNonreciprocalLandauZener2022,ivakhnenkoNonadiabaticLandauZener2023}, occurring when the wave packets pass through anti-crossing levels at $p=0$.

From Hamiltonian (\ref{eq:1}) and Bloch oscillation solution (\ref{eq:5}), we can derive the Hamiltonian for the Landau-Zener tunneling in the periodically driven two-level quantum system \cite{shevchenkoLandauZenerStuckelberg2010}:
\begin{equation}\label{eq:6}
    H_\mathrm{LZ}=\begin{pmatrix}
        M & 2\pi\alpha p_c(t) \\2\pi\alpha p_c(t)&-M
    \end{pmatrix}.
\end{equation}
The eigenvalues of $H_\mathrm{LZ}$ at each momentum site $p$ are $E_p^\pm=\pm\sqrt{(2\pi\alpha p)^2+M^2}$, and the corresponding eigenstates are:
\begin{equation}\label{eq:7}
    \lvert\pm\rangle_p\propto
    \begin{pmatrix}M\pm\sqrt{(2\pi\alpha p)^2+M^2}\\2\pi\alpha p\end{pmatrix}.
\end{equation}
The occupation probability of each energy level is given by $P_\pm=\sum_{p\in\mathbb{Z}}\lvert\psi_p^\pm\rvert^2$, where $\lvert\psi_p^\pm\rvert^2$ is the occupation probability at each momentum site $p$.

As shown in Fig.~\ref{fig:3}, the occupation probability transitions between the two eigenenergy levels as the wave packet undergoes Landau-Zener tunneling while passing through the anti-crossing levels, driven by the kicks.

\begin{figure}[t]
    \includegraphics[width=0.45\textwidth]{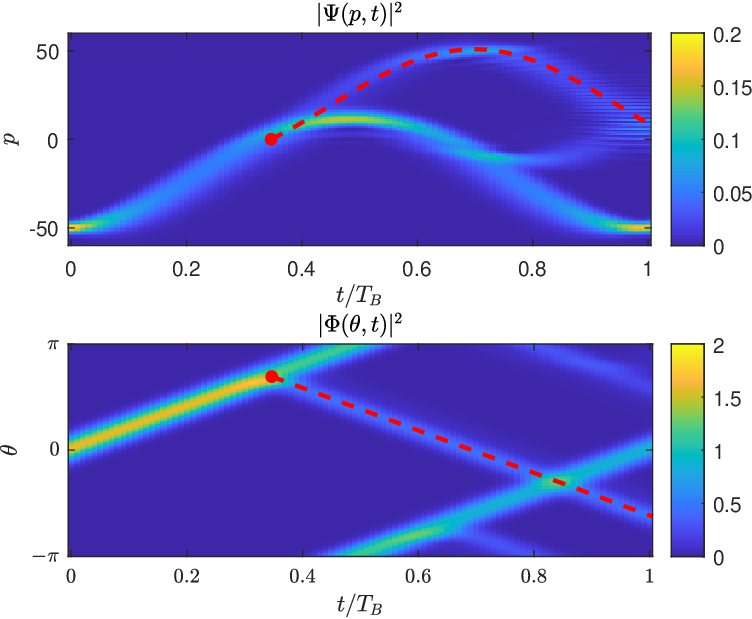}
    \caption{Wave packet splitting in both momentum and coordinate spaces. The red dot denotes the splitting point, while the red dashed lines indicate the subsequent trajectories of the new branches. All parameters are identical to those in Fig.~\ref{fig:1}(c).}
    \label{fig:4}
\end{figure}

The splitting observed in Figs.~\ref{fig:1}(b)(c) at $p=0$ can be explained by this two-level Landau-Zener tunneling process, where part of the wave packet tunnels to the other eigenenergy level, preserving the direction of oscillation. The remaining part alters the sign of angular coordinate growth, i.e., the sign in Eq.~(\ref{eq:2}). As shown in Fig.~\ref{fig:4}, the splitting becomes more pronounced in the angular coordinate representation as the wave packet passes through the anti-crossing levels.
The time at which the wave packet crosses the $p=0$ line is obtained from Eq.~(\ref{eq:5}):
\begin{equation}\label{eq:8}
    t_1=\frac{1}{2\pi\alpha}\arccos\left(\frac{2\pi\alpha p_0}{A}+\cos\delta_s\right)-\frac{\delta_s}{2\pi\alpha},\quad\in(0,T_B).
\end{equation}
The corresponding coordinate and momentum at $t_1$ are $\theta_1=\theta_0+2\pi\alpha t_1$ and $p_1=0$. The part of the wave packet that remains in the original eigenenergy level will use $(p_1,\theta_1)$ as the starting point for subsequent Bloch oscillations in the opposite direction. As shown in Fig.~\ref{fig:4}, the red dots represent $(t_1,p_1)$ and $(t_1,\theta_1)$, and the red dashed lines represent the subsequent dynamical evolution after $t_1$, which align well with the observed splitting branches.
\begin{figure}[t]
    \includegraphics[width=0.45\textwidth]{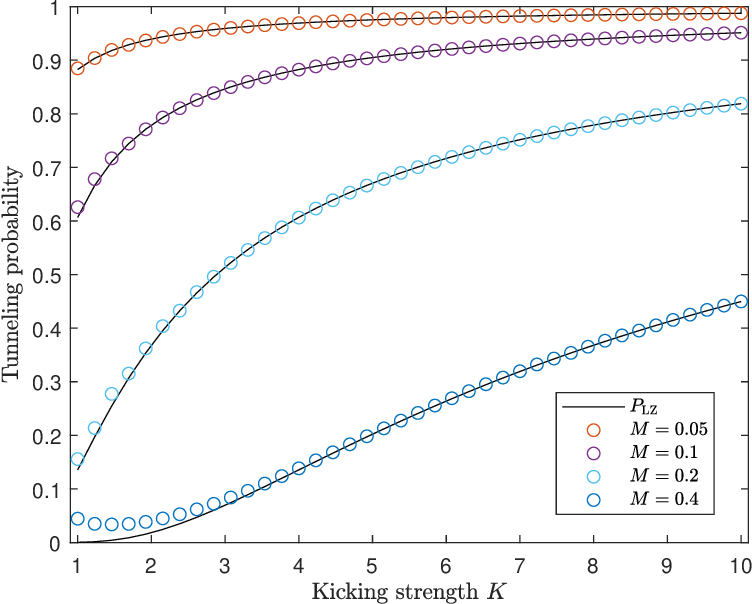}
    \caption{The single passage Landau-Zener tunneling probability. The black lines denote the theoretical prediction $P_\mathrm{LZ}$ given by Eq.~(\ref{eq:10}), while the colored circles correspond to numerical simulations for various mass $M$. All other parameters are identical to those in Fig.~\ref{fig:1}.}
    \label{fig:5}
\end{figure}

We linearize $p_c(t)$ around $t_1$ with the initial condition $(p_0,\theta_0=0)$:
\begin{equation}\label{eq:9}
    p_c(t)=A\sin(2\pi\alpha t_1+\delta_s)(t-t_1)+\mathcal{O}(t-t_1)^2.
\end{equation}
Here, $A\sin(2\pi\alpha t_1+\delta_s)\equiv v$ represents the velocity of the wave packet center crossing the anti-crossing at $p=0$. And $2\pi\alpha v$ corresponds to the sweeping rate of the Landau-Zener tunneling process.

Therefore, we can calculate the single-passage Landau-Zener tunneling probability \cite{shevchenkoLandauZenerStuckelberg2010} as:
\begin{equation}\label{eq:10}
    \begin{split}
        P_\mathrm{LZ}
         & =\exp\left[-2\mathrm{Im}\int^\tau_0E_+(t)-E_-(t)\mathrm{d}t\right]                                  \\
         & =\exp\left(-\frac{M^2}{2\alpha\sqrt{\left\lvert A^2-(2\pi\alpha p_0+A\cos\delta_s)^2\right\rvert}}\right).
    \end{split}
\end{equation}
Here, $E_\pm(t)=\pm\sqrt{M^2+(2\pi\alpha vt)^2}$ represents the linearized energy levels of $H_\mathrm{LZ}$ near $t_1$, and $\tau=i\frac{M}{2\pi\alpha v}$ denotes the complex zero of $E_\pm(t)$.

As shown in Fig.~\ref{fig:5}, we validate this Landau-Zener formula by comparing it with our numerical simulations over various kicking strengths with various masses. The wave packet initialized with spin state $\frac{1}{\sqrt{2}}[1,1]^T$, sitting at lower energy level $E_-$. We set initial momentum as $p_0=-\frac{K}{3\pi\alpha}$ to ensure tunneling occurs. The results show great agreement between the theoretical prediction and numerical simulations. The observed deviation arises because the initial spin state $\chi$ is not perfectly aligned with the eigenstates of $H_\mathrm{LZ}$ at $p_0$, particularly for large $M$ and small $K$.
\subsection{From BLZ dynamics to dynamical localization: relativistic and spin effects}\label{sec:3D}
\begin{figure}
    \includegraphics[width=0.45\textwidth]{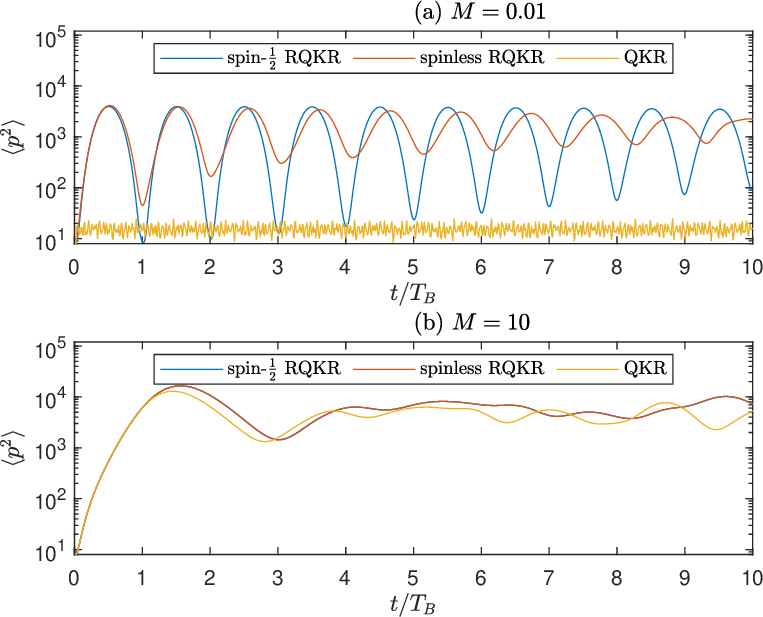}
    \caption{Time evolution of the momentum spread $\langle p^2\rangle$ for a Gaussian wave packet. The dynamical evolution is governed by these three Hamiltonians Eqs.~(\ref{eq:1}), (\ref{eq:11}) and (\ref{eq:12}), for various mass values $M$.}
    \label{fig:6}
\end{figure}
In the previous subsections, we have thoroughly investigated the Bloch oscillation and Landau-Zener tunneling of the spin-$\frac{1}{2}$ RQKR. While, the conventional QKR system exhibits the prominent phenomenon of dynamical localization in momentum space. To investigate how the relativistic effects and spin degrees of freedom affect the dynamical evolution is also of interest.

We first consider the positive energy branch (electron branch) of spin-$\frac{1}{2}$ RQKR. The spinful model reduces to a spinless RQKR, governed by the Hamiltonian:
\begin{equation}\label{eq:11}
    H_\mathrm{spinless~RQKR}=\sqrt{(2\pi\alpha p)^2+M^2}+K\cos\theta\sum_{n=-\infty}^{+\infty}\delta(t-nT).
\end{equation}
In the non-relativistic limit $(2\pi\alpha p)/M\rightarrow0$, this spinless RQKR further reduces to the conventional QKR:
\begin{equation}\label{eq:12}
    H_\mathrm{QKR}=\frac{2\pi^2\alpha^2}{M}p^2+M+K\cos\theta\sum_{n=-\infty}^{+\infty}\delta(t-nT).
\end{equation}

Numerical simulations of the time evolution of momentum diffusion $\langle p^2 \rangle$, governed by these three Hamiltonians Eqs.~(\ref{eq:1}), (\ref{eq:11}), and (\ref{eq:12}), for various mass values $M$ are presented in Fig.~\ref{fig:6}. The wavepackets are initially centered at $p_0=0$, and initial spin state $\chi=[1,0]^T$ is chosen for spin-$\frac{1}{2}$ QRKR, corresponding to the positive energy branch.

As shown in Fig.~\ref{fig:6}(a), for a relatively small mass ($M=0.01$), the time evolution of momentum spread $\langle p^2\rangle$ in both the spin-$\frac{1}{2}$ and spinless RQKR models exhibit distinct oscillatory behavior, with a period roughly equals to our theoretically predicted $T_B=1/\alpha$ in the first few periods. The oscillation amplitudes decay, and periods change gradually in the long-term evolution. Compared to the spinless RQKR model, the spin-$\frac{1}{2}$ one sustains the periodic oscillations for a longer time duration. In contrast, for the conventional QKR, $\langle p^2\rangle$ shows prominent localization behavior without any discernible oscillation.

When increasing the particle mass (i.e., $M=10$) in Fig.~\ref{fig:6}(b), both the spin-$\frac{1}{2}$ and spinless RQKR models approach the behavior of the conventional QKR, i.e., $\langle p^2\rangle$ show a saturation behavior after a short-term growth, and no discernible oscillatory behavior observed in all of these three models.

When the particle mass is small, the energy expression of $\sqrt{(2\pi\alpha p)^2+M^2}$ degenerates to be an approximately linear dependence on momentum $p$. A relatively small mass will imply a strong relativistic effect, which supports Bloch oscillations, as discussed in Sec.~\ref{sec:3B}. The decay of the Bloch oscillation is due to the small quadratic $p^2$ modification on the energy expression, which can induce the dynamical localization effects~\cite{fishmanChaosQuantumRecurrences1982,casatiStochasticBehaviorClassical1979,chirikovQuantumChaosLocalization1988}, analogous to the disorder inhibited Bloch oscillation in solid state physics~\cite{schulteDynamicsBlochOscillations2008,drenkelBlochOscillationsBoseEinstein2008}. When spin degrees of freedom are included, tunneling between two energy levels ($\pm\sqrt{(2\pi\alpha p)^2+M^2}$) becomes possible. The tunneled wavepackets moving along the asymptotic line of the avoid level crossing equivalently have the energy of linear dependence on momentum $p$, which helps sustain the oscillatory behavior over a relatively longer time duration as shown in Fig.~\ref{fig:6}(a).

A large particle mass implies a weak relativistic effect. In this case, the $p^2$ term becomes predominant in the energy expression as shown in Eq.~(\ref{eq:12}). The systems' dynamics in this regime resemble the conventional QKR model, where the dynamical localization effects totally inhibit the relativistic oscillatory behavior as shown in Fig.~\ref{fig:6}(b).
\section{conclusion}\label{sec:4}
In conclusion, we have investigated the momentum-space dynamics of a periodically kicked Dirac particle numerically and analytically. Using the Floquet formalism, we uncover the Bloch oscillations and Landau-Zener tunneling underlying the dynamical behavior of spin-$\frac{1}{2}$ RQKR. These Bloch-Landau-Zener dynamics can be regarded as a momentum-space counterpart to a tilted bipartite lattice system. The BLZ dynamics offer an interpretive framework for spin dynamics described in Ref.~\cite{rozenbaumDynamicalLocalizationCoupled2017} and connects to recent studies on similar dynamics in other systems \cite{keBlochLandauZenerDynamicsSingleparticle2015,nederBlochOscillationsLandau2024,rahmanBlochOscillationPhases2024,pratesBlochLandauZenerOscillationsQuasiperiodic2024}. Additionally, we have discussed how relativistic effects and spin degrees of freedom influence the dynamical properties in these periodically kicked quantum rotor systems. These findings could provide valuable insights for future experimental studies.
\begin{acknowledgments}
    This research was supported by the National Science Foundation of China (Grant No. U2330401). We thank Dr.~Fulong Dong, Dr.~Shiwei Liu, and Dr.~Bingbin Wu for their helpful advice.
\end{acknowledgments}
\bibliographystyle{apsrev4-2}

\begin{thebibliography}{55}
    \bibitem{abaninRecentProgressManybody2017}
    D.~A. Abanin and Z.~Papi\'c, Annalen der Physik \textbf{529}, 1700169 (2017).
    \bibitem{abaninColloquiumManybodyLocalization2019}
    D.~A. Abanin, E.~Altman, I.~Bloch, and M.~Serbyn, Rev. Mod. Phys. \textbf{91}, 021001 (2019).
    \bibitem{rudnerBandStructureEngineering2020}
    M.~S. Rudner and N.~H. Lindner, Nat. Rev. Phys. \textbf{2}, 229 (2020).
    \bibitem{harperTopologyBrokenSymmetry2020}
    F.~Harper, R.~Roy, M.~S. Rudner, and S.~Sondhi, Annu. Rev. Condens. Matter Phys. \textbf{11}, 345 (2020).
    \bibitem{izrailevSimpleModelsQuantum1990}
    F.~M. Izrailev, Physics Reports \textbf{196}, 299 (1990).
    \bibitem{saifClassicalQuantumChaos2005}
    F.~Saif, Physics Reports \textbf{419}, 207 (2005).
    \bibitem{santhanamQuantumKickedRotor2022}
    M.~Santhanam, S.~Paul, and J.~B. Kannan, Physics Reports \textbf{956}, 1 (2022).
    \bibitem{casatiStochasticBehaviorClassical1979}
    G.~Casati, J.~Ford, J.~Ehlers, K.~Hepp, R.~Kippenhahn, H.~A. Weidenm\"uller, J.~Zittartz, and W.~Beiglb\"ock, eds., \emph{Stochastic Behavior in Classical and Quantum Hamiltonian Systems}, Lecture Notes in Physics, Vol.~{93} (Springer, Berlin, 1979) pp. {334--352}.
    \bibitem{izrailevQuantumResonanceRotator1980}
    F.~M. Izrailev and D.~L. Shepelyanskii, Theor. Math. Phys. \textbf{43, 553} (1980).
    \bibitem{chirikovQuantumChaosLocalization1988}
    B.~Chirikov, F.~Izrailev, and D.~Shepelyansky, Physica D: Nonlinear Phenomena \textbf{33}, 77 (1988).
    \bibitem{chirikovUniversalInstabilityManydimensional1979}
    B.~V. Chirikov, Physics Reports \textbf{52}, 263 (1979).
    \bibitem{mooreAtomOpticsRealization1995}F.~L. Moore, J.~C. Robinson, C.~F. Bharucha, B.~Sundaram, and M.~G. Raizen, Phys. Rev. Lett. \textbf{75}, 4598 (1995).
    \bibitem{ammannQuantumDeltaKickedRotor1998}
    H.~Ammann, R.~Gray, I.~Shvarchuck, and N.~Christensen, Phys. Rev. Lett. \textbf{80}, 4111 (1998).
    \bibitem{andersonAbsenceDiffusionCertain1958}
    P.~W. Anderson, Phys. Rev. \textbf{109}, 1492 (1958).
    \bibitem{fishmanChaosQuantumRecurrences1982}
    S.~Fishman, D.~R. Grempel, and R.~E. Prange, Phys. Rev. Lett. \textbf{49}, 509 (1982).
    \bibitem{russomannoChaosSubdiffusionInfiniterange2021}
    A.~Russomanno, M.~Fava, and R.~Fazio, Phys. Rev. B \textbf{103}, 224301 (2021).
    \bibitem{zhaoQuantizationOutoftimeorderedCorrelators2022}
    W.-L. Zhao, Phys. Rev. Research \textbf{4}, 023004 (2022).
    \bibitem{zhaoScalingLawsOutoftimeorder2023}
    W.-L. Zhao, R.-R. Wang, H.~Ke, and J.~Liu, Phys. Rev. A \textbf{107}, 062201 (2023).
    \bibitem{tomsovicControllingQuantumChaos2023}
    S.~Tomsovic, J.~D. Urbina, and K.~Richter, Phys. Rev. E \textbf{108}, 044202 (2023).
    \bibitem{zhaoPhaseModulationDirected2024}
    W.-L. Zhao, G.~Li, and J.~Liu, Phys. Rev. Research \textbf{6}, 033249 (2024).
    \bibitem{chernikovChaosRelativisticGeneralization1989}
    A.~A. Chernikov, T.~T\'el, G.~Vattay, and G.~M. Zaslavsky, Phys. Rev. A \textbf{40}, 4072 (1989).
    \bibitem{kimRelativisticChaosDriven1995}J.-H.
    Kim and H.-W. Lee, Phys. Rev. E \textbf{51}, 1579 (1995).
    \bibitem{huangRelativisticQuantumChaos2018}
    L.~Huang, H.-Y. Xu, C.~Grebogi, and Y.-C. Lai, Physics Reports \textbf{753}, 1 (2018).
    \bibitem{matrasulovRelativisticKickedRotor2005}
    D.~U. Matrasulov, G.~M. Milibaeva, U.~R. Salomov, and B.~Sundaram, Phys. Rev. E \textbf{72}, 016213 (2005).
    \bibitem{zhaoQuantumClassicalSuperballistic2014}
    Q.~Zhao, C.~A. M\" uller, and J.~Gong, Phys. Rev. E \textbf{90}, 022921 (2014).
    \bibitem{rozenbaumDynamicalLocalizationCoupled2017}
    E.~B. Rozenbaum and V.~Galitski, Phys. Rev. B \textbf{95}, 064303 (2017).
    \bibitem{gerritsmaQuantumSimulationDirac2010}
    R.~Gerritsma, G.~Kirchmair, F.~Z\"ahringer, E.~Solano, R.~Blatt, and C.~F. Roos, Nature \textbf{463}, 68 (2010).
    \bibitem{zhangRelativisticQuantumEffects2012}
    D.-w. Zhang, Z.-d. Wang, and S.-l. Zhu, Front. Phys. \textbf{7}, 31 (2012).
    \bibitem{silvaOpticalSimulationFree2019}
    T.~L. Silva, E.~R.~F. Taillebois, R.~M. Gomes, S.~P. Walborn, and A.~T. Avelar, Phys. Rev. A \textbf{99}, 022332 (2019).
    \bibitem{longhiKleinTunneling}
    S.~Longhi, Phys. Rev. B \textbf{81}, 075102 (2010).
    \bibitem{dreisowBlochZenerOscillations}
    F.~Dreisow, A.~Szameit, M.~Heinrich, T.~Pertsch, S.~Nolte, A.~T\"unnermann, S.~Longhi, Phys. Rev. Lett. \textbf{102}, 076802 (2009).
    \bibitem{wilczekTimeCrystal}
    M.~P.~Zaletel, M.~Lukin, C.~Monroe, C.~Nayak, F.~Wilczek, and N.~Y.~Yao, Rev. Mod. Phys. \textbf{95}, 031001 (2023).
    \bibitem{grempelLocalizationIncommensuratePotential1982}
    D.~R. Grempel, S.~Fishman, and R.~E. Prange, Phys. Rev. Lett. \textbf{49}, 833 (1982).
    \bibitem{longhiMarylandModelOptical2021}
    S.~Longhi, Opt. Lett. \textbf{46}, 637 (2021).
    \bibitem{prangeSolvableModelQuantum1984}
    R.~E. Prange, D.~R. Grempel, and S.~Fishman, Phys. Rev. B \textbf{29}, 6500 (1984).
    \bibitem{berryIncommensurabilityExactlysolubleQuantal1984}
    M.~Berry, Physica D: Nonlinear Phenomena \textbf{10}, 369 (1984).
    \bibitem{simonAlmostPeriodicSchrodinger1985}
    B.~Simon, Annals of Physics \textbf{159}, 157 (1985).
    \bibitem{shirleySolutionSchrodingerEquation1965}
    J.~H. Shirley, Phys. Rev. \textbf{138}, B979 (1965).
    \bibitem{yangEffectsPeriodicallyKicked2024}
    F.~Yang, Z.~Wei, T.-M. Li, and S.-P. Kou, Phys. Rev. B \textbf{109}, 195123 (2024).
    \bibitem{blochUeberQuantenmechanikElektronen1929}
    F.~Bloch, Z. Physik \textbf{52}, 555 (1929).
    \bibitem{zenerTheoryElectricalBreakdown1934}
    C.~Zener, Proc. R. Soc. Lond. A \textbf{145}, 523 (1934).
    \bibitem{hartmannDynamicsBlochOscillations2004}
    T.~Hartmann, F.~Keck, H.~J. Korsch, and S.~Mossmann, New J. Phys. \textbf{6}, 2 (2004).
    \bibitem{breidBlochZenerOscillations2006}
    B.~M. Breid, D.~Witthaut, and H.~J. Korsch, New J. Phys. \textbf{8}, 110 (2006).
    \bibitem{flossAndersonWallBloch2014}
    J.~Flo\ss and I.~S. Averbukh, Phys. Rev. Lett. \textbf{113}, 043002 (2014).
    \bibitem{landauZurTheorieEnergieubertragung1932}
    L.~D. Landau, Physikalische Zeitschrift der Sowjetunion \textbf{2} (1932).
    \bibitem{zenerNonadiabaticCrossingEnergy1932}
    C.~Zener, Proc. R. Soc. Lond. A \textbf{137}, 696 (1932).
    \bibitem{grifoniDrivenQuantumTunneling1998}
    M.~Grifoni and P.~H\"anggi, Physics Reports \textbf{304}, 229 (1998).
    \bibitem{liuTheoryNonlinearLandauZener2002}
    J.~Liu, L.~Fu, B.-Y. Ou, S.-G. Chen, D.-I. Choi, B.~Wu, and Q.~Niu, Phys. Rev. A \textbf{66}, 023404 (2002).
    \bibitem{shevchenkoLandauZenerStuckelberg2010}
    S.~Shevchenko, S.~Ashhab, and F.~Nori, Physics Reports \textbf{492}, 1 (2010).
    \bibitem{wangAdiabaticityNonreciprocalLandauZener2022}
    W.-Y. Wang, B.~Sun, and J.~Liu, Phys. Rev. A \textbf{106}, 063708 (2022).
    \bibitem{ivakhnenkoNonadiabaticLandauZener2023}
    O.~V. Ivakhnenko, S.~N. Shevchenko, and F.~Nori, Physics Reports \textbf{995}, 1 (2023).
    \bibitem{schulteDynamicsBlochOscillations2008}
    Schulte, T., S. Drenkelforth, G. Kleine Büning, W. Ertmer, J. Arlt, M. Lewenstein, and L. Santos. Phys. Rev. A \textbf{77}, 023610 (2008).
    \bibitem{drenkelBlochOscillationsBoseEinstein2008}
    Drenkelforth, S, G Kleine Büning, J Will, T Schulte, N Murray, W Ertmer, L Santos, and J J Arlt. New J. Phys. \textbf{10}, 045027 (2008).
    \bibitem{keBlochLandauZenerDynamicsSingleparticle2015}
    Y.~Ke, X.~Qin, H.~Zhong, J.~Huang, C.~He, and C.~Lee, Phys. Rev. A \textbf{91}, 053409 (2015).
    \bibitem{nederBlochOscillationsLandau2024}
    I.~Neder, C.~Sirote-Katz, M.~Geva, Y.~Lahini, R.~Ilan, and Y.~Shokef, Proc. Natl. Acad. Sci. U.S.A. \textbf{121}, e2310715121 (2024).
    \bibitem{rahmanBlochOscillationPhases2024}
    T.~Rahman, A.~Wirth-Singh, A.~Ivanov, D.~Gochnauer, E.~Hough, and S.~Gupta, Phys. Rev. Research \textbf{6}, L022012 (2024).
    \bibitem{pratesBlochLandauZenerOscillationsQuasiperiodic2024}
    H.~C. Prates and V.~V. Konotop, Phys. Rev. Research \textbf{6}, L022011 (2024).
\end{thebibliography}

\end{document}